\documentclass{llncs}
\usepackage{amssymb}
\usepackage{graphicx}
\usepackage{cite}
\usepackage{color}
\usepackage{textcomp}
\usepackage{url}
\usepackage{caption}
\usepackage{siunitx}
\usepackage{tabularx}

\usepackage{academicons}
\usepackage{hyperref}
\usepackage{xcolor}
\newcommand{\orcid}[1]{\href{https://orcid.org/#1}{\includegraphics[scale=0.08]{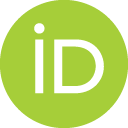}}}

\usepackage{url}
\urldef{\mailsa}\path|{alfred.hofmann, ursula.barth, ingrid.haas, frank.holzwarth,|
\urldef{\mailsb}\path|anna.kramer, leonie.kunz, christine.reiss, nicole.sator,|
\urldef{\mailsc}\path|erika.siebert-cole, peter.strasser, lncs}@springer.com|    
\newcommand{\keywords}[1]{\par\addvspace\baselineskip
\noindent\keywordname\enspace\ignorespaces#1}

\hypersetup{colorlinks=false}

\begin{document}

\mainmatter  

\title{Long-range complex in the HC$_{3}$N + CN potential energy surface: ab initio calculations and intermolecular potential}

\author{Emília Valença Ferreira de Aragão\inst{1,2}\orcid{0000-0002-8067-0914} 
\and Luca Mancini\inst{2}
\and Noelia Faginas-Lago\inst{2}\orcid{0000-0002-4056-3364} 
\and Marzio Rosi\inst{3}\orcid{0000-0002-1264-3877} 
\and  Nadia Balucani\inst{2}\orcid{0000-0001-5121-5683}
\and  Fernando Pirani\inst{2}\orcid{0000-0003-3110-6521}
}
\authorrunning{E.V.F. Aragão et al.}

\institute{
Master-tec srl, Via Sicilia 41, 06128 Perugia, Italy\\
\email{emilia.dearagao@master-tec.it}\\
\and
Dipartimento di Chimica, Biologia e Biotecnologie,\\ Universit\`{a} degli Studi di Perugia, 06123 Perugia, Italy\\ 
\email{\{emilia.dearagao,luca.mancini2\}@studenti.unipg.it}\\
\email{\{noelia.faginaslago,nadia.balucani,fernando.pirani\}@unipg.it}\\
\and
 Dipartimento di Ingegneria Civile ed Ambientale,\\ Universit\`{a} degli Studi di Perugia, 06125 Perugia, Italy\\
 \email{marzio.rosi@unipg.it}
}

\toctitle{Lecture Notes in Computer Science}
\tocauthor{Authors' Instructions}
\maketitle

\begin{abstract}

In this work we characterize an initial van der Waals adduct in the potential energy surface of reaction between cyanoacetylene HC$_{3}$N and the cyano radical.
The geometry of the CN-HC$_{3}$N adduct has been optimized through calculations employing ab initio methods. 
Results show that the energy of the adduct lays below the reactants.
Additionally, a saddle point that connects the adduct to an important intermediate of the PES has been been localized, with energy below the reactants.
Calculations of the intermolecular potential have been performed and results show that the energy of the van der Waals adduct is higher than estimated with the ab initio methods.

\end{abstract}

\section{Introduction}

To this date, more than 200 molecular species have been detected in the interstellar medium (ISM)~\cite{mcguire2018survey}.
Among the first observed organic molecules is HC$_{3}$N, or cyanoacetylene, identified in the early 1970’s by Turner~\cite{turner1971detection} towards the galactic radio source SgrB2.
In the same decade, new observations have been reported: in 1976~\cite{morris1976cyanoacetylene} cyanoacetylene has been detected in 17 galactic sources, including dark clouds and a molecular envelope around a carbon star, while later in 1979 Walmsley reported the detection of HC$_3$N in TMC2~\cite{walmsley1980cyanoacetylene}.
Later, the detection of  HC$_{5}$N, HC$_{7}$N, HC$_{9}$N and HC$_{11}$N has also been reported~\cite{broten1978Adetection,bell1997detection}.
These molecules share the same characteristics: they are linear, with alternating carbon-carbon single and triple bonds and have a cyano group.
As a whole, this group of molecules is addressed as cyanopolyynes and the members of this group have been detected in a number of environments in the interstellar medium: hot cores, star forming regions, cold clouds and even solar-type protostars~\cite{wyrowski1999vibrationally,takano1998observations,mendoza2018search, taniguchi2018survey,jaber2017history}.
In addition, HC$_{3}$N has also been detected in low-mass protostars~\cite{bergner2017complex}, protoplanetary disks~\cite{bergner2018survey} and on Titan's atmosphere~\cite{kunde1981hc3ntitan}.

It is currently accepted that the synthesis of cyanopolyynes in the ISM is a result of successive chain-elongation reactions involving cyanoacetylene and C$_{2}$H radical~\cite{cheikh2013low}.
However, when cyanoacetylene collides with a CN radical, the reaction might lead to the formation of dicyanoacetylene (NCCCCN). 
This prevents elongation reactions from occurring, therefore the reaction with cyano radical is considered as a chain-termination reaction.
Knowing how effective this reaction is and how fast it proceeds is important to estimate the distribution of nitrogen in the insterstellar medium.
Yet, the product of the reaction, dicyanoacetylene, is not detectable in the ISM due to the lack of permanent electric dipole moment.
Therefore, we need to estimate the rate of formation of this molecule through experiments and using computational tools. 

A previous study by Petrie and Osamura~\cite{petrie2004nccn} reports the results of their investigation on the HNC + C$_{3}$N potential energy surface (PES), which share some points in common with the HC$_{3}$N + CN PES.
At that time they had optimized the geometries at B3LYP/6-311G** level and computed single-point energies at CCSD(T)/aug-cc-pVDZ level.
They have reported a channel where, from HNC + C$_{3}$N, the system has to overcome an energy barrier to form a linear van der Waals adduct, before dissociating into HC$_{3}$N and CN.
In addition, another channel shows the formation of NCCCCN from HC$_{3}$N and CN, but no van der Waals adduct has been reported for this channel.
In this work, we report a connection between the linear van der Waals and the intermediate of the channel that forms dicyanoacetylene. 

In relation to the kinetics of this reaction, an experimental value and a theoretical estimate of the rate coefficient have been reported by Cheikh et al~\cite{cheikh2013low}.
In particular, the theoretical estimate was performed employing a two transition-state model, which implies the formation of a van der Waals adduct.
We initially thought that the linear van der Waals adduct could be involved in the estimate of the rate coefficient. 
However, as it will be shown, another approach between the fragments must be considered.
As it will be discussed, the ab initio approach allows us to locate only one of the van der Waals complexes, while the evaluation of the intermolecular potential shows the possibility of more. 
In particular, the Improved Lennard-Jones (ILJ)\cite{pirani2008beyond} function will allow a better description of the long-range interaction within the intermolecular potential.
A correct estimation of the weak intermolecular interactions is a key point for an accurate estimate of a reaction  rate constant. The ILJ potential can be considered as a refinement of the classical Lennard-Jones potential, which represents the potential as a function of the distance between the two interacting fragments. In particular the ILJ function gives a better reproduction  of both the long-range attraction and the short-range repulsion with a simple formulation. 

In the present work we investigate the initial approach of the cyanoacetylene and cyano radical at two specific angles. 
We try to understand how the most stable intermediate of the potential energy surface is formed.
We employ two different theoretical methods to identify the formation of a van der Waals adduct.
Identifying this type of structure is important for the estimation of the rate constant.

\section{Methods}

In this section we present two approaches we have employed in this work: the ab initio methods and the semi-empirical method.

\subsection{Ab Initio Calculations}

The open-shell system HC$_{3}$N-CN has been investigated following a well established computational strategy~\cite{depetris2011proton,leonori2009observation,depetris2007ssoh,rosi2013theoretical,skouteris2019interstellar,sleimanPCCP2018,berteloite2011low}.
Such strategy consists in performing electronic structure calculations for the reactants, intermediates and saddle point on the doublet potential energy surface (PES).
Geometry optimizations and harmonic vibrational frequency analysis have been carried out employing density functional theory (DFT), with the Becke-3-parameter exchange and Lee-Yang-Parr correlation (B3LYP)\cite{becke1993density,stephens1994ab}. 
Harmonic frequency analysis serves to estimate the zero-point energy correction and to confirm if an optimized structure is a stationary point. In that case, an optimized geometry is assigned as a minimum if all the frequencies are real and as a saddle point if one single frequency is imaginary.
For saddle points, Intrinsic Reaction Coordinate (IRC) calculations \cite{gonzalez1989improved,gonzalez1990reaction} have also been performed.
Frequency analysis were followed by a single-point calculation performed employing coupled-cluster, including single and double excitations as well as perturbative estimate of connected triples (CCSD(T))\cite{bartlett1981many,raghavachari1989fifth,olsen1996full}.
The resulting computed energy is combined with the zero-point correction from the frequency calculation to correct it to 0 K.
Both B3LYP and CCSD(T) methods have been used along with the correlation consistent valence polarized basis set aug-cc-pVTZ\cite{dunning1989gaussian}.
All electronic structure calculations were performed using the Gaussian09 software\cite{frisch2009gaussian}.
Data on the geometry was collected using the open-source software Avogadro\cite{hanwell2012avogadro}.

\subsection{Intermolecular potential}
%
%
In addition to the ab initio calculations, a semi-empirical study has been done on the evolution of the intermolecular interaction energy as the two reactants approach each other in a specific angle.
In the adopted approach, each interacting partner is assumed to be composed by \textit{effective} atoms distributed on the molecular frame. 
Accordingly, the total intermolecular potential \textit{V$_{total}$} has been defined as combination of a non-electrostatic and an electrostatic component:
\begin{equation}
V_{total}=V_{non-elect} + V_{elect}
\label{eq:1}
\end{equation}
\noindent
%
Moreover, each component has been represented as additivity of different atom-atom interaction pair contributions. 
In particular, each non-electrostatic atom-atom contribution has been provided by the ILJ function, formulated as
\begin{equation}V_{ILJ}(r) =
\varepsilon \left[\frac{m}{n(r)-m}\left (
\frac{r_{m}}{r} \right)^{n(r)} -
\frac{n(r)}{n(r)-m}\left(
\frac{r_{m}}{r}\right)^m \right]\label{eq:2}
\end{equation}
\noindent
where $\varepsilon$ is the depth of the potential well, $r_{m}$ is its position and $r$ is the atom-atom distance.
The parameter $n(r)$ was firstly introduced in a 2004 paper by Pirani et al.~\cite{pirani04:37} as a modification of the Maitland-Smith correction to the Lennard-Jones model~\cite{MAITLAND1973443}.
According to the authors, $m$ assumes the value of 6 for neutral-neutral systems, 4 for ion-induced dipole, 2 for ion-permanent dipole, and 1 for ion-ion cases.

Eq.~\ref{eq:2} is composed by a repulsive term and a term that represents the long-range attraction, both expressed in function of $r$.
To modulate the decline of the repulsion and the strength of the attraction, $n(r)$ takes the following form:
\begin{equation}
n(r) = \beta +4.0\left(\frac{r}{r_{m}}\right)^2\label{eq:3}
\end{equation}
\noindent
where $\beta$ is a parameter attributed to the nature and the hardness of the interacting particles.~\cite{pirani04:37,pacifici13:2668,bart2008,cappelletti2008}. In the current systems, the value assigned to $\beta$ is 7. The values for $\varepsilon$ and $r_{m}$ parameters, specific to each atom-atom pair, are reported in Table~\ref{tab1}.
%
They have been obtained exploiting the \textit{effective} polarizability of various interacting atoms, whose combinations are consistent with the average polarizability of two molecular partners. 
%
\begin{table}[t]
\centering
\caption{Parameters for the non-electrostatic potential.\label{tab1}}
\begin{tabularx}{\linewidth}
{ | >{\centering\arraybackslash}X 
  | >{\centering\arraybackslash}X 
  | >{\centering\arraybackslash}X |}
\hline
Interacting pair  & r$_{m}$ (\si{\angstrom}) & $\varepsilon$ (meV)  \\ 
\hline
C-C$_{Rad}$      & 3.80       &  6.16\\ 
N-C$_{Rad}$     & 3.73       &  6.07\\ 
H-C$_{Rad}$      & 3.54       &  2.43\\ 
C-N$_{Rad}$      & 3.71       &  5.95\\ 
N-N$_{Rad}$     & 3.63       &  6.07\\ 
H-N$_{Rad}$      & 3.41       &  2.55\\ 
\hline
\end{tabularx}
\end{table}
\begin{table}[h]
\centering
\caption{Parameters for the electrostatic potential.\label{tab2}}
\begin{tabularx}{\linewidth}
{ | >{\centering\arraybackslash}X 
  | >{\centering\arraybackslash}X| }
\hline
Atom  & Mulliken partial charges (CCSD(T))\\ 
\hline
C$_{1}$           &  -1.720\\ 
C$_{2}$           &  1.762\\ 
C$_{3}$           &  0.088\\ 
N                 &  -0.583\\ 
H                 &  0.453\\ 
 &                       \\
C$_{Rad}$         &  0.152\\ 
N$_{Rad}$         &  -0.152\\ 
 \hline
\end{tabularx}
\end{table}
Within the same approach, the electrostatic component \textit{V$_{elect}$} is expressed as
\begin{equation}
V_{elect}(r) = \frac{1}{4\pi\varepsilon_{0}} \sum_{i=1}^{5}\sum_{j=1}^{2}\frac{q_iq_j}{r_{ij}}\label{eq:4}
\end{equation}
\noindent
where $\varepsilon_{0}$ is the vacuum permittivity, \textit{q$_{i}$} corresponds to the partial charge on an atom of the cyanoacetylene molecule, \textit{q$_{j}$} is the partial charge on an atom of the cyano radical and \textit{r$_{ij}$} is the distance between the two atoms involved.
In this work, the electrostatic contribution has been computed with charge values extracted from the CCSD(T)/aug-cc-pVTZ single-point calculations of the separated reactants.
Those values are reported on Table~\ref{tab2}.

\section{Results}
In this section we present the results of the ab initio calculations and show how the intermolecular potential can provide additional insight into the formation of the van der Waals complex.
\subsection{Ab initio Calculations}
\begin{figure}[h]
\centering
\includegraphics[width=\linewidth]{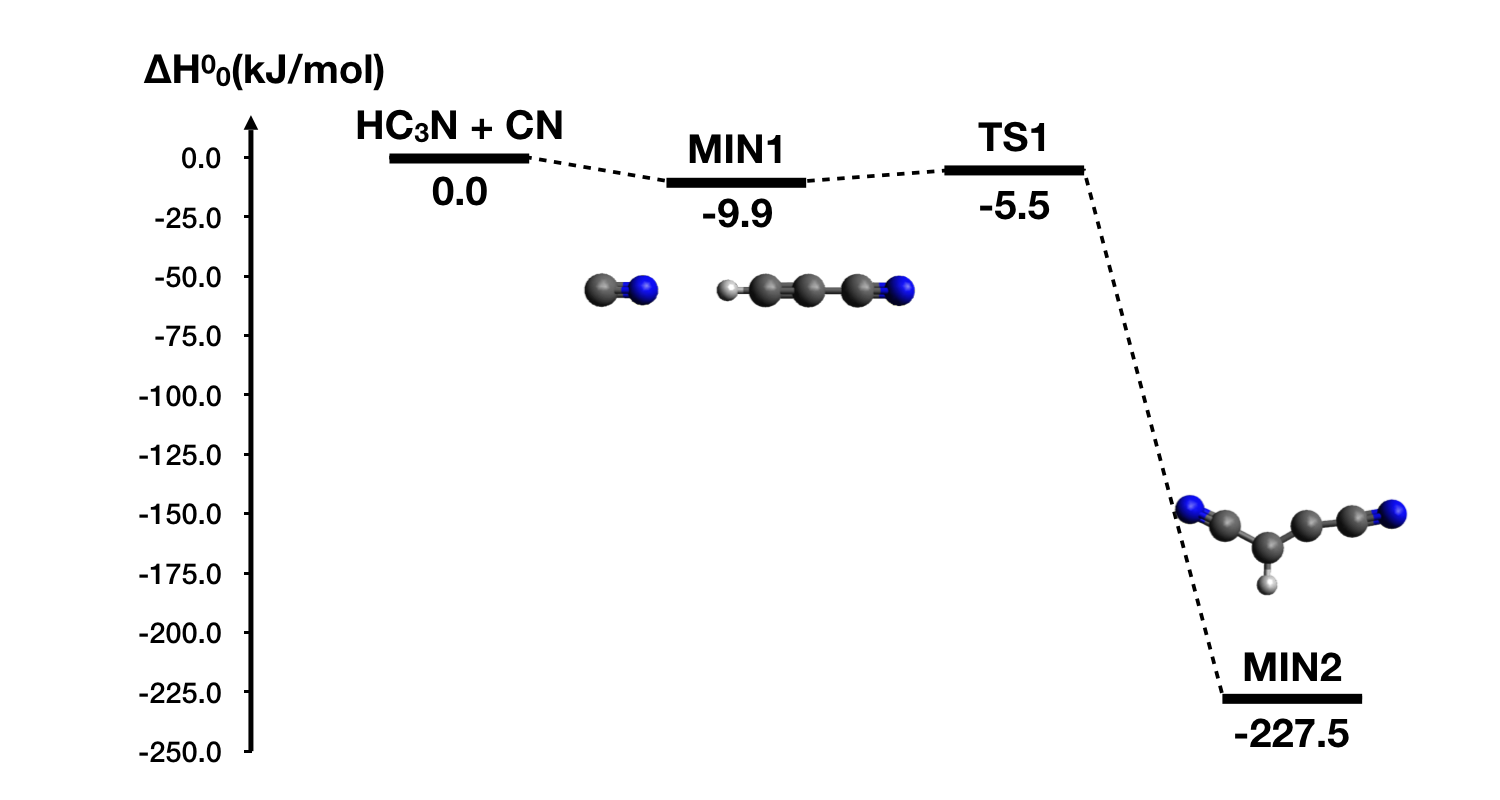}
\caption{Potential energy surface of the cyanoacetylene (HC$_{3}$N) and cyano radical (CN) reaction. Energies computed at CCSD(T)/aug-cc-pVTZ level with zero-point corrections at B3LYP/aug-cc-pVTZ. }
\label {fig1}
\end{figure}
The data obtained from electronic structure calculations with the ab initio method shows the formation of a long-range complex as first step of the reaction between cyanoacetylene and cyano radical.
Figure~\ref{fig1} reports a potential energy surface depicting the formation of a bonded intermediate, with energy values relative to the reactants.
Four points of the PES were characterized: reactants, a van der Waals complex (MIN1), a saddle point (TS1) and the bonded intermediate (MIN2).
The process starts with a barrierless formation of the van der Waals complex MIN1.
The system must then overcome a barrier through TS1 to form the bonded intermediate MIN2, in which we can notice the formation of chemical bond between carbon atoms.
This process is exothermic at CCSD(T)/aug-cc-pVTZ level of calculation considering geometries optimized at B3LYP/aug-cc-pVTZ level.

\begin{figure}[t]
\centering
\includegraphics[width=\linewidth]{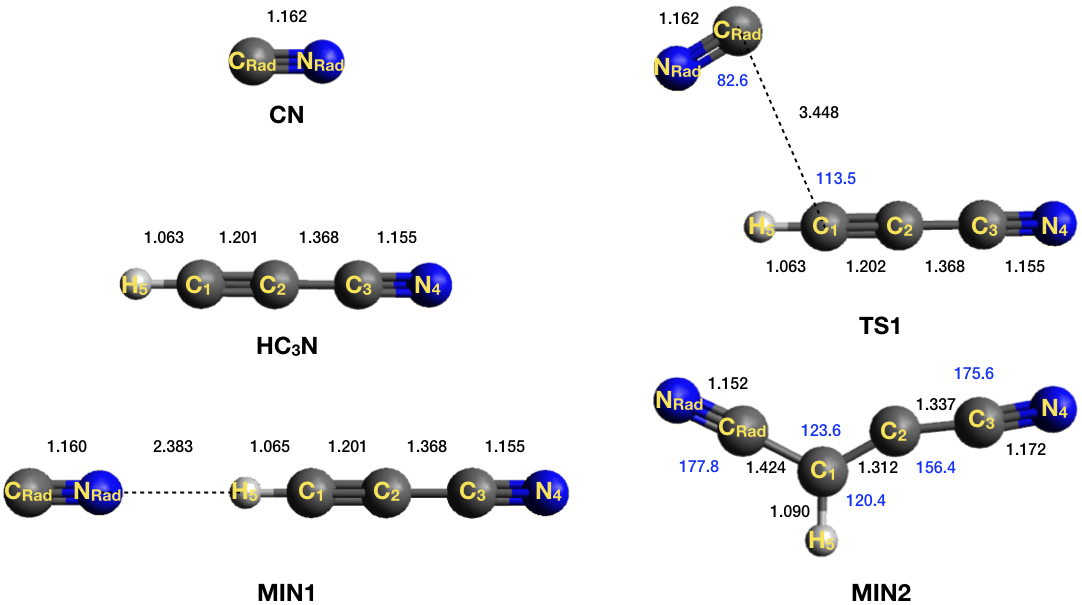}
\caption{Structural details of the model representation of cyanoacetylene (HC$_{3}$N), cyano radical (CN), the van der Waals complex (MIN1), the saddle point (TS1) and the bonded intermediate (MIN2) geometries optimized at the B3LYP/aug-cc-pVTZ level. Bond lengths (black) are shown in \si{\angstrom}, while bond angles (blue) are displayed in degrees. Carbon atoms are represented in grey, nitrogen atoms in blue and hydrogen atoms in white.}
\label {fig2}
\end{figure}
Figure~\ref{fig2} depicts the geometry of the reactants, the van der Waals complex (MIN1), the saddle point (TS1) and the bonded intermediate (MIN2) with labeled atoms, distances in angstroms and angles in degrees.
The structure of MIN1 shows the interaction between the hydrogen atom of cyanoacetylene and the nitrogen atom of the radical.
As it can be observed, the formation of the complex does not promote significant changes in the bond distances.
On the other hand, the transition state TS1 shows a very different geometry: the CN fragment appears to rotate in order to form the carbon-carbon bond between C$_{1}$ and C$_{Rad}$.
At last, MIN2 shows the formed carbon-carbon bond and a loss of linearity between the atoms from cyanoacetylene.

\subsection{Intermolecular potential}

\begin{figure}[t]
\centering
\includegraphics[width=\linewidth]{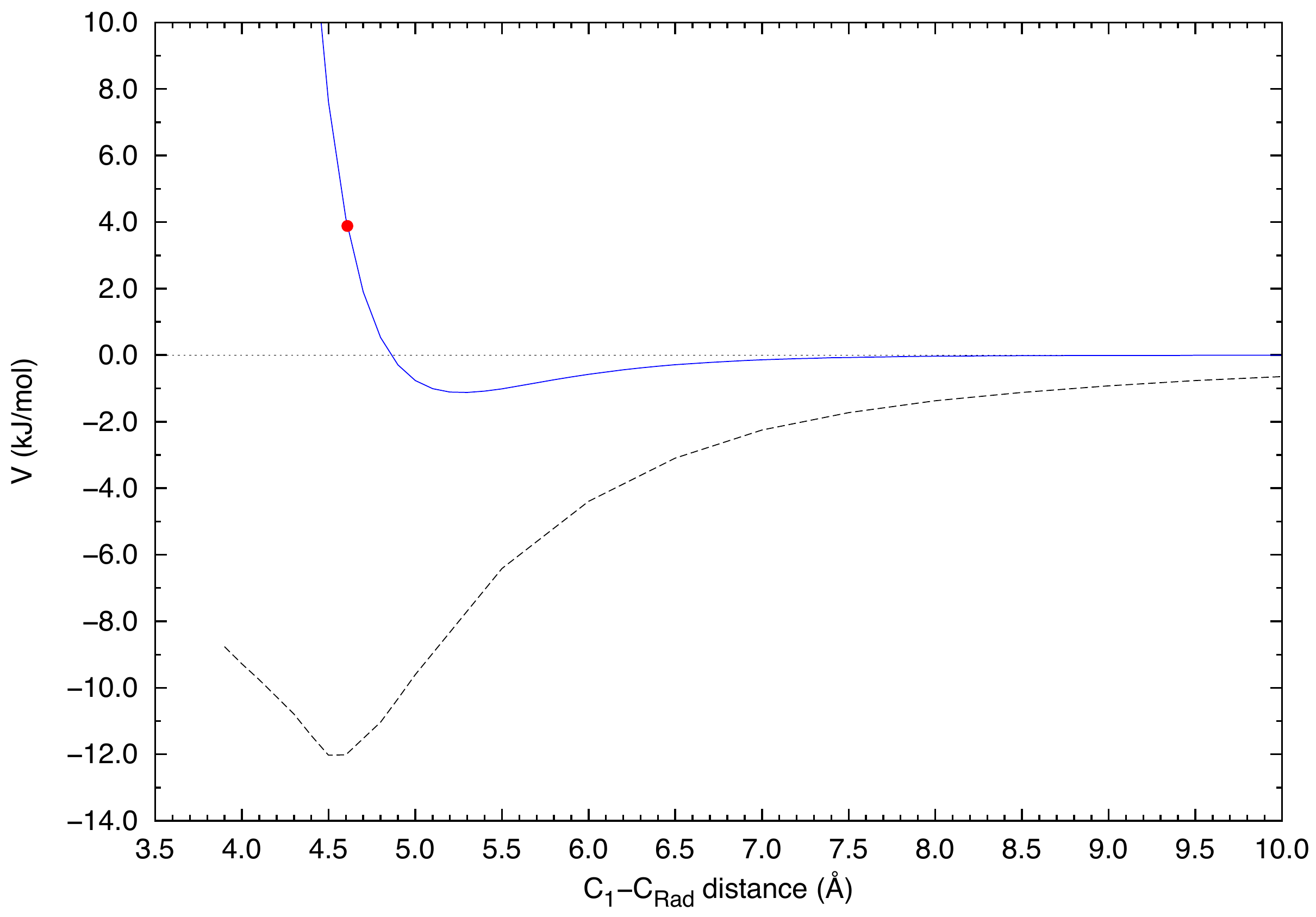}
\caption{Curve of the intermolecular potential(blue,full line) and curve of the variational calculation (black, dashed) corresponding to the linear approach between fragments. Energies of the variational curve were computed at CCSD(T)/aug-cc-pVTZ level. The intermolecular energy of the van der Waals complex MIN1 is also represented (red dot).}
\label {fig3}
\end{figure}
\begin{figure}[t]
\centering
\includegraphics[width=\linewidth]{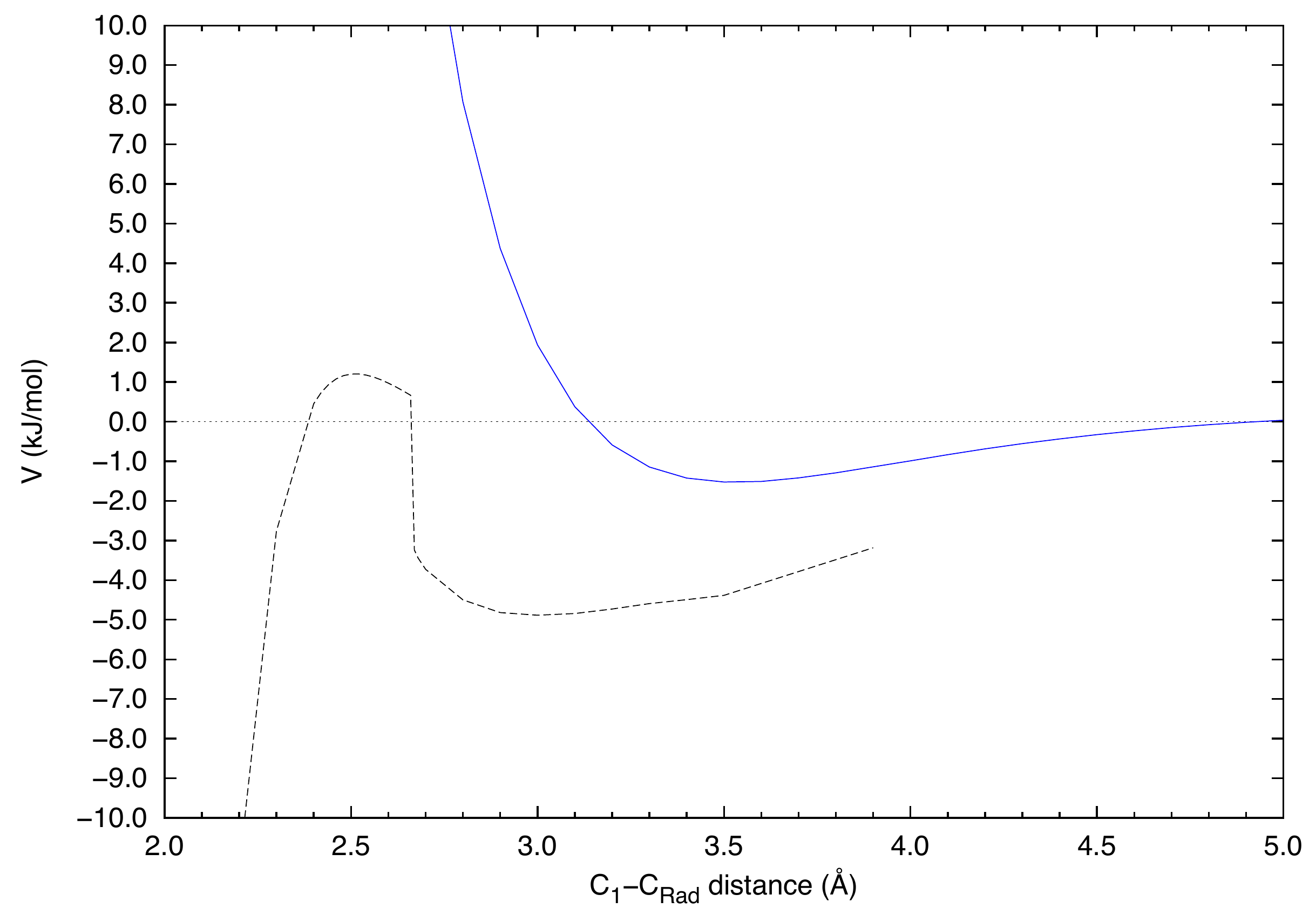}
\caption{Curve of the intermolecular potential (blue,full line) and curve of the variational calculation (black, dashed) corresponding to the orthogonal approach between fragments. Energies of the variational curve were computed at CCSD(T)/aug-cc-pVTZ level.}
\label {fig4}
\end{figure}
\begin{figure}[t]
\centering
\includegraphics[width=\linewidth]{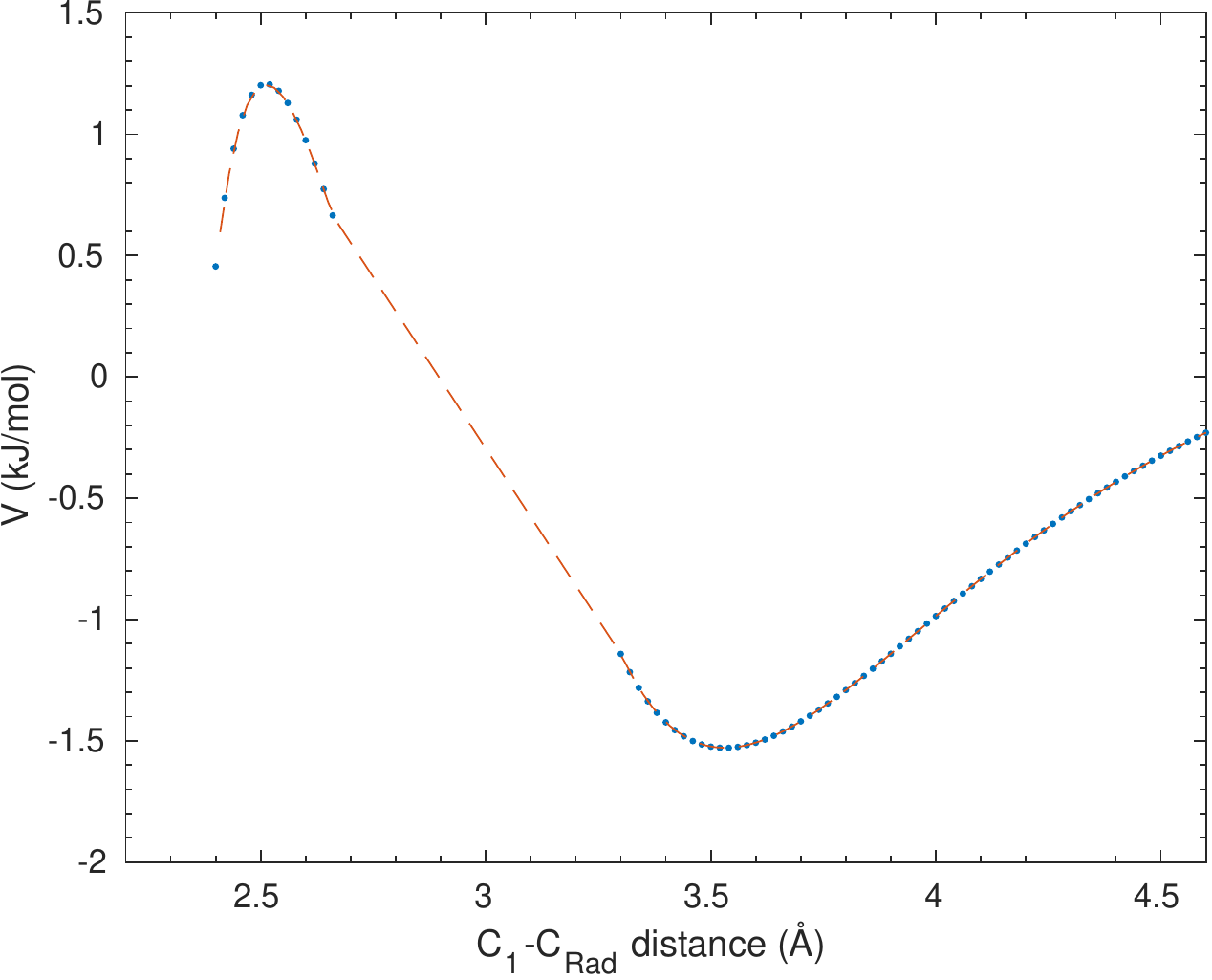}
\caption{Linear interpolation (red,dashed) between the curve of the intermolecular potential and curve of the variational calculation corresponding to the orthogonal approach between fragments. The points of the original curves are represented as blue dots.}
\label {fig5}
\end{figure}
As it has been discussed in the methods section, the intermolecular potential expression can be decomposed into repulsion and attraction contribution terms.
In this section we compare the evolution of the intermolecular potential with variational calculations performed with the ab initio methods when the two reactants approach each other.

In Figure~\ref{fig3} are displayed the curve of the intermolecular potential in blue and the curve of the variational calculation in black.
In the x-axis we chose to represent the distance between carbons C$_{1}$ and C$_{Rad}$ since it is the most relevant distance involved in the formation of intermediate MIN2.
The curve of the intermolecular potential was 
built by taking the separate geometries of the reactants and approaching them in a linear way to form MIN1.
In every point of the curve, the atom-atom pair distances were calculated and were employed to estimate the non-electrostatic and the electrostatic potential energies using the parameters in tables~\ref{tab1} and ~\ref{tab2} respectively.
The sum of both terms results in a total intermolecular energy in meV, then converted into kJ/mol after multiplying it by a factor of 0.096487.
The variational curve is the results of the variational calculation.
Each point of the curve corresponds to an optimized geometry at a fixed C$_{1}$-C$_{Rad}$ distance.
This was possible by using the keyword "Modredundant" in the G09 code.
The energies reported in the graph were obtained at CCSD(T)/aug-cc-pVTZ level and were not corrected at 0K.
Here, the energy of the separated reactants is set to zero and all other energies are relative to this energy.

Three aspects of this graph must be discussed.
First, at long range, the values of the ab initio calculations are underestimated. They are expected to be closer to zero in a similar manner as the interatomic potential.
This implies that the long-range interaction is not well described by this approach.
Second, the well in both curves are located at different distances and have different depths.
For the intermolecular potential, the point of the lowest energy value was located at 5.3 \si{\angstrom} and had a correspondent V = -1.123 kJ/mol.
For the curve of the variational calculation, the geometry with the lowest energy was close to the MIN1 geometry, or at \textit{d$_{C_{1}-C_{Rad}}$} = 4.5 \si{\angstrom} and V = -12.03 kJ/mol.
If we take the geometry of MIN1 and compute it with the intermolecular potential (red dot on the blue curve), the correspondent energy value is +3.88 kJ/mol.
This means that relative to the intermolecular potential, the ab initio calculation seems to be underestimating the effects of the repulsion.
At last, in the intermolecular potential curve, the energy grows exponentially for distances shorter than 4.9 \si{\angstrom}, due to the repulsion.
In the variational calculation, for distances shorter than 4.4 \si{\angstrom} we notice a change in the fragments approach. The fragments starts to rotate in order to form the MIN2, and most points in this part of the curve shows a geometry closer to TS1.

Figure~\ref{fig4}, following the same scheme as figure~\ref{fig3}, corresponds to an attack of 90° degrees between fragments.
Contrary to the linear approach, the corresponding van der Waals adduct could not be located through calculations employing ab initio methods.
Attempts to find this geometry usually ended with the formation of MIN2.
However, the curve corresponding to the variational approach shows the formation of a minimum around 3.0 \si{\angstrom}. 
In a similar manner, the intermolecular potential curve shows the existence of a minimum around 3.5 \si{\angstrom}, with V = -1.52 kJ/mol.
At shorter distances, around 2.5 \si{\angstrom}, the variational curve shows the presence of a transition state, which also could not be located with ab initio calculations.
The subsequent decrease in energy corresponds to the formation of the MIN2.
The intermolecular curve shows only an increase in energy since, contrary to the variational calculations, the reactants geometries are frozen.

In figure~\ref{fig5}, a linear interpolation between the two curves was made considering values from 2.40 to 2.66 \si{\angstrom} of the variational curve and values from 3.3 and 5.2 \si{\angstrom} of the intermolecular potential curve.
The resulting graph represents the formation of the van der Waals adduct from an attack at 90° degrees and the presence of a barrier that must be overcome to form MIN2. 

\section{Conclusions}

Though a linear van der Waals adduct has been optimized through ab initio calculations, we cannot conclude with ab initio method alone that this corresponds to the path of formation of the most stable intermediate i.e MIN2.
The approach of the two fragments at an angle of 90° degrees was taken into consideration. 
The calculations performed employing the semi-empirical method showed that another van der Waals structure can be localised at 3.5 \si{\angstrom}.
Variational calculations, which employed ab initio methods, shows the presence of a barrier around 2.5 \si{\angstrom}.
These results characterise better the formation of MIN2.

\section{Acknowledgements}

This project has received funding from the European Union’s Horizon
2020 research and innovation programme under the Marie Sk{\l}odowska Curie grant agreement No 811312 for the project ”Astro-Chemical Origins” (ACO).  
E. V. F. A and N.F-L. thanks the Herla Project (http://hscw.herla.unipg.it) - Universit\`{a} degli Studi di Perugia for allocated computing time. 
The authors thank the Dipartimento di Ingegneria Civile e Ambientale of the University of Perugia for allocated computing time within the project “Dipartimenti di Eccellenza 2018-2022”.
N. F.-L thanks MIUR and the University of Perugia for the financial support of the AMIS project through the “Dipartimenti di Eccellenza” programme.
N. F.-L also acknowledges the Fondo Ricerca di Base 2020
(RICBASE2020FAGINAS) del Dipartimento di Chimica, Biologia e Biotecnologie della Università di Perugia for financial support. 
We thank J. Vekeman, I. García Cuesta and A. Sánchez de Merás for provinding the codes to carry out the fittings.
The authors thank S. Meniconi and C. Capponi of the University of Perugia for assistance with the MATLAB code.

\bibliography{EVFA_ICCSA2021}{}

\bibliographystyle{splncs04} 
\end{document}